\documentclass{ws-mpla}
\newcounter{comments}
\renewcommand{\thecomments}{C\arabic{comments}}

\begin{document}

\title{Neutrinoless double-beta decay. A brief review}

\author{S. M. BILENKY}

\address{Bogoliubov Laboratory of Theoretical Physics, Joint Institute for Nuclear Research,\\
Dubna, R-141980, Russia;\\INFN, Sezione di Torino, Via P. Giuria 1, I--10125 Torino, Italy}

\author{CARLO GIUNTI}

\address{INFN, Sezione di Torino, Via P. Giuria 1, I--10125 Torino, Italy}

\maketitle


\begin{abstract}
In this brief review we discuss the
generation of Majorana neutrino masses through the see-saw mechanism,
the theory of neutrinoless double-beta decay,
the implications of neutrino oscillation data for the effective Majorana mass,
taking into account the recent Daya Bay
measurement of $\vartheta_{13}$,
and the interpretation of the results of
neutrinoless double-beta decay experiments.
\end{abstract}

\ccode{PACS Nos.: 14.60.Pq, 14.60.Lm, 14.60.St}

\section{Introduction}
\label{intro}

One of the most important recent discoveries in particle physics is
the observation of neutrino oscillations in atmospheric\cite{1002.3471},
solar\cite{1109.0763}, reactor\cite{1009.4771} and accelerator
\cite{hep-ex/0212007,1103.0340} neutrino experiments. Neutrino oscillations is a
quantum-mechanical consequence of the neutrino mixing relation
\begin{equation}\label{mixing}
\nu_{lL}(x) = \sum^{3}_{i=1} U_{li} \, \nu_{iL}(x) \qquad (l=e,\mu,\tau)
.
\end{equation}
Here $\nu_{i}(x)$ is the field of neutrinos with mass $m_{i}$, $U$ is
the 3$\times$3 unitary PMNS\cite{Pontecorvo:1957cp,Pontecorvo:1958qd,Maki:1962mu} mixing matrix. The
left-handed flavor field $\nu_{lL}(x)$ enters into the standard
leptonic charged current
\begin{equation}\label{lepCC}
j^{CC}_{\alpha}(x) = 2 \sum_{l=e,\mu,\tau} \bar{\nu}_{lL}(x) \, \gamma_{\alpha} \, l_L(x)
\end{equation}
and determines the notion of a left-handed flavor neutrino
$\nu_{l}$ which is produced in CC weak processes together with a lepton
$l^{+}$. The flavor neutrino $\nu_{l}$ is described by the mixed
state
\begin{equation}\label{state}
|\nu_{l}\rangle=\sum_{i=1}^{3} U^{*}_{li} \, |\nu_{i}\rangle
,
\end{equation}
where $|\nu_{i}\rangle $ is the state of a neutrino with mass $m_{i}$
and a definite momentum.

The probability of the transition $\nu_{l}\to \nu_{l'}$ in vacuum is
given by the standard expression (see Ref.\cite{hep-ph/9812360})
\begin{equation}\label{probability}
P(\nu_{l}\to\nu_{l'})=
\left|
\sum_{i}
U_{l'i}
\,
e^{-iE_{i}t}
\,
U^{*}_{li}
\right|^2
=
\left|\delta_{l'l}+
\sum_{i\neq k}U_{l'i} \left(e^{-i\frac{\Delta{m}_{ki}^2L}{2E}}-1\right) U^{*}_{li} \right|^2
.
\end{equation}
Here $\Delta{m}_{ki}^2= m_{i}^2- m_{k}^2$, $L \simeq t$ is the distance between the neutrino
detector and the neutrino source, and $E$ is the neutrino energy.

In the standard parameterization,
the $3\times 3$ PMNS mixing matrix is characterized by three mixing
angles, $\vartheta_{12}$, $\vartheta_{23}$ and $\vartheta_{13}$, by a Dirac CP-violating phase
$\delta$
and by two possible Majorana CP-violating phases $\lambda_{2}$ and $\lambda_{3}$:
\begin{equation}
U
=
\begin{pmatrix}
c_{12}
c_{13}
&
s_{12}
c_{13}
&
s_{13}
e^{-i\delta}
\\
-
s_{12}
c_{23}
-
c_{12}
s_{23}
s_{13}
e^{i\delta}
&
c_{12}
c_{23}
-
s_{12}
s_{23}
s_{13}
e^{i\delta}
&
s_{23}
c_{13}
\\
s_{12}
s_{23}
-
c_{12}
c_{23}
s_{13}
e^{i\delta}
&
-
c_{12}
s_{23}
-
s_{12}
c_{23}
s_{13}
e^{i\delta}
&
c_{23}
c_{13}
\end{pmatrix}
D(\lambda_{2},\lambda_{3})
\,,
\label{MixMat}
\end{equation}
where
$ c_{ab} \equiv \cos\vartheta_{ab} $
and
$ s_{ab} \equiv \sin\vartheta_{ab} $.
The diagonal matrix
$
D(\lambda_{2},\lambda_{3})
=
\mathrm{diag}(1,e^{i\lambda_{2}},e^{i\lambda_{3}})
$
is present only if massive neutrinos are Majorana particles.
The Majorana phases
have an effect in processes
which are allowed only if massive neutrinos are Majorana particles
and are characterized by a violation of the total lepton number,
as neutrinoless double-beta decay
(see Section~\ref{effmass}).
Since neutrino oscillations are flavor transitions without violation of the total lepton number,
they do not depend on the Majorana
phases\cite{Bilenky:1980cx,Doi:1980yb,Langacker:1986jv,1001.0760}.
The neutrino oscillation probabilities depend only on
the four mixing parameters
$\vartheta_{12}$, $\vartheta_{23}$, $\vartheta_{13}$ and $\delta$,
and on two independent mass-squared differences
$\Delta{m}_{12}^2$ and $\Delta{m}_{23}^2$.
From the analysis of the
experimental data it follows that
\begin{equation}\label{inequality}
\Delta{m}_{12}^2\simeq \frac{1}{30} \, |\Delta{m}_{23}^2|
.
\end{equation}
In the case of three-neutrino mixing
assumed in Eqs.~(\ref{mixing}) and (\ref{state}),
two neutrino mass spectra
are possible:
\begin{enumerate}
\item Normal spectrum (NS)
\begin{equation}\label{norspec}
m_{1}< m_{2} < m_{3} ;\quad \Delta{m}^2_{12} \ll \Delta
m^2_{23}
.
\end{equation}
\item Inverted spectrum (IS)
\begin{equation}\label{invspec}
m_{3}< m_{1} < m_{2} ;\quad \Delta{m}^2_{12} \ll
| \Delta{m}^2_{13}|
.
\end{equation}
\end{enumerate}
The existing experimental data do not allow to establish, what type of
neutrino mass spectrum is realized in nature.

Let us introduce the "solar" and "atmospheric" mass-squared differences
$\Delta{m}^2_{s}$ and $\Delta{m}^2_{a}$, respectively. For both
spectra we have $\Delta{m}^2_{12}=\Delta{m}^2_{s}$. For normal
(inverted) spectrum we have $\Delta{m}^2_{23}=\Delta{m}^2_{a}$ ($|\Delta{m}^2_{13}|=\Delta{m}^2_{a}$).

From a three-neutrino analysis of the Super-Kamiokande data\cite{1002.3471},
the values of the neutrino oscillation parameters in the case of a normal
(inverted) mass spectrum are, at 90\% C.L.,
\begin{eqnarray}\label{SKanalysis}
&
1.9 \, (1.7)\cdot 10^{-3} \, \mathrm{eV}^2 \leq\Delta{m}_{a}^2 \leq
2.6 \, (2.7)\cdot 10^{-3} \, \mathrm{eV}^2,
&
\nonumber
\\
&
0.407 \leq \sin^2\vartheta_{23} \leq 0.583,\quad \sin^2\vartheta_{13} <
0.04 \, (0.09)
.
&
\end{eqnarray}
The results of the Super-Kamiokande atmospheric neutrino experiment
have been fully confirmed by the long-baseline accelerator neutrino experiments
K2K\cite{hep-ex/0212007} and MINOS\cite{1103.0340}.

From the two-neutrino analysis of the MINOS
$\nu_{\mu}\to \nu_{\mu}$ data, for the parameters
$\Delta{m}_{a}^2$ and $\sin^22\vartheta_{23}$ the following values were obtained:
\begin{equation}\label{Minos1}
\Delta{m}_{a}^2=(2.32 {}_{-0.08}^{+0.12})\cdot 10^{-3} \, \mathrm{eV}^2,\quad \sin^22\vartheta_{23}>0.90
.
\end{equation}

From the combined three-neutrino analysis of all solar neutrino data and the data
of the reactor KamLAND experiment, it was found that\cite{1009.4771}
\begin{equation}\label{solkam}
\Delta{m}_{s}^2=(7.50 {}_{-0.20}^{+0.19})\cdot 10^{-5} \, \mathrm{eV}^2,
\quad
\tan^2\vartheta_{12}=
0.452{}^{+0.035}_{-0.033},
\quad
\sin^2\vartheta_{13}=0.020\pm 0.016
.
\end{equation}
From a similar analysis performed by the SNO collaboration, it was
obtained that\cite{1109.0763}
\begin{equation}\label{solkam1}
\Delta{m}_{s}^2=(7.41 {}_{-0.19}^{+0.21})\cdot 10^{-5} \, \mathrm{eV}^2,\quad \tan^2\vartheta_{12}=
0.446{}^{+0.030}_{-0.029},\quad \sin^2\vartheta_{13}=0.025 {}^{+0.018}_{-0.015}
.
\end{equation}

The Daya Bay collaboration\cite{1203.1669} measured recently with high precision the mixing angle
$\vartheta_{13}$:
\begin{equation}
\sin^2\vartheta_{13}
=
0.024 \pm 0.004
.
\label{s13db}
\end{equation}
This is a $5.2\sigma$ evidence of a non-zero value of $\vartheta_{13}$
which confirms the previous measurements of
T2K\cite{1106.2822},
MINOS\cite{1108.0015} and
Double Chooz\cite{1112.6353}.
It also confirms earlier indications of a non-zero value of $\vartheta_{13}$
found in the analysis of the data of solar and other neutrino experiments
(see Eqs.~(\ref{solkam}) and (\ref{solkam1}) and Ref.\cite{0809.2936,0804.3345,0806.2649,0810.5443,1001.4524}).
The Daya Bay measurement has important implications for theory\cite{1203.1672} and experiment
(see Ref.\cite{1003.5800}).
It opens promising perspectives for the observation of CP violation in the lepton sector
and matter effects in long-baseline experiments,
which could allow to determine the character of the neutrino mass spectrum.

Several years ago
an indication in favor of
short-baseline $\bar\nu_{\mu}\to \bar\nu_{e}$ transitions
was found in the LSND experiment
\cite{hep-ex/0104049}.
The LSND data can be explained by neutrino oscillations
with $0.2 \, \mathrm{eV}<\Delta{m}^2< 2 \, \mathrm{eV}$ and $10^{-3}<
\sin^2 2\vartheta<4\cdot 10^{-2}$. Recently an additional (2$\sigma$)
indication in favor of short-baseline oscillations, compatible with
the LSND result, was obtained in the MiniBooNE experiment\cite{1007.1150}.
Moreover,
the data obtained in old reactor short-baseline experiments
can also be interpreted as indications in favor of oscillations\cite{1101.2755} by using
a new calculation of the reactor neutrino fluxes\cite{1101.2663,1106.0687}.
All these data (if confirmed) imply that the number of massive
neutrinos is larger than three and in addition to the three flavor neutrinos
$\nu_{e},\nu_{\mu},\nu_{\tau}$ mixed sterile neutrinos
$\nu_{s_{1}},...$ must exist.

The problem of short-baseline neutrino oscillations and sterile
neutrinos is a hot topic at the moment. Several new short-baseline
reactor and accelerator experiments are aimed to check this
possibility in the near future (see ref.\cite{1012.4356}).

The absolute values of neutrino masses are currently unknown.
The Mainz\cite{hep-ex/0412056} and Troitsk\cite{1108.5034} experiments on
the high-precision measurement of the end-point part of the
$\beta$-spectrum of $^{3}H$ decay
found the 95\% C.L. upper bounds
\begin{equation}\label{MainzTr}
m_{\beta} \leq 2.3 \, \mathrm{eV}
\, (\mathrm{Mainz}),
\qquad
m_{\beta} \leq 2.1 \, \mathrm{eV}
\, (\mathrm{Troitsk}),
\end{equation}
for the ``average'' neutrino mass
(see Ref.\cite{Giunti-Kim-2007})
\begin{equation}
m_{\beta}=\sqrt{\sum_{i}|U_{ei}|^2m^2_{i}}
\,.
\label{trimass}
\end{equation}

From neutrino oscillation and tritium $\beta$-decay data we conclude
that
\begin{enumerate}
\renewcommand{\labelenumi}{(\theenumi)}
\renewcommand{\theenumi}{\Alph{enumi}}
\item Neutrino masses are different from zero.
\item Neutrino masses are much smaller than the masses of charged leptons and quarks.
\item Neutrino masses are not (or not only) of Standard Model (SM) Higgs origin.
\end{enumerate}

Several mechanisms of neutrino mass generation have been proposed.
It is widely believed that the most plausible one is the seesaw mechanism\cite{Minkowski:1977sc,GellMann-Ramond-Slansky-SeeSaw-1979,Yanagida-SeeSaw-1979,Mohapatra:1980ia}.
According to this mechanism, small neutrino masses are
generated by new interactions beyond the SM which violates the total
lepton number $L$ at a scale much larger than the electroweak scale
$v=(\sqrt{2}G_{F})^{-1/2}\simeq 246$ GeV.

If the seesaw mechanism is realized, {\em the neutrinos $\nu_{i}$ with definite
masses are Majorana particles} and, consequently, the lepton
number violating neutrinoless double-beta decay
($0\nu\beta\beta$-decay) of even-even nuclei,
\begin{equation}\label{betabeta}
N(A,Z) \to N(A,Z+2) +e^{-} +e^{-},
\end{equation}
is allowed,
where $N(A,Z)$ is a nucleus with nucleon number $A$ and proton number $Z$.
The knowledge of the nature of neutrinos with definite masses (Majorana
or Dirac?) is extremely important for the understanding of the origin of
small neutrino masses.
Using large detector masses, high energy
resolutions and low backgrounds, the experiments on the search for
neutrinoless double-beta decay allow to reach unparalleled sensitivities
to extremely small effects due to the Majorana neutrino masses. In this brief
review we consider this process (see also
\cite{Bilenky:2002aw,Elliott:2002xe,Elliott:2004hr,0708.1033,1001.1946,1106.1334,GomezCadenas:2011it,Schwingenheuer:2012jt}).

\section{Seesaw mechanism of neutrino mass generation}
\label{seesaw}

In this Section we briefly discuss the standard seesaw mechanism of neutrino mass
generation\cite{Minkowski:1977sc,GellMann-Ramond-Slansky-SeeSaw-1979,Yanagida-SeeSaw-1979,Mohapatra:1980ia}.
We consider a general approach based on the effective
Lagrangian formalism\cite{Weinberg:1979sa}. Let us assume that
the Standard Model is valid up to some
scale $\Lambda$.
If we include effects of physics beyond the SM,
the total Lagrangian (in the SM region)
has the form
\begin{equation}\label{EffL}
\mathcal{L}(\Lambda)= \mathcal{L}^{SM}+\sum_{n\geq 1}
\frac{1}{\Lambda^{n}}\mathcal{O}_{4+n}
.
\end{equation}
The second term is a nonrenormalizable part of the Lagrangian.
It is built from SM fields and satisfies the requirement of
$SU(2)\times U(1)$ invariance. The operator $\mathcal{O}_{4+n}$
has dimension $M^{4+n}$.

In the expansion (\ref{EffL}) of the non-renormalizable part of the Lagrangian in powers of $1/\Lambda$,
the most
important term for neutrino physics is the first one,
$\mathcal{L}^{\rm{eff}}_{I} = \mathcal{O}_{5} / \Lambda$,
which contains an operator of dimension five.
This term can be built
from the leptons and Higgs doublets:
\begin{equation}\label{effL1}
\mathcal{L}^{\rm{eff}}_{I}
=
-
\frac{1}{\Lambda}
\sum_{l',l}
\left[
\overline L_{l'L} \tilde{H}
\right]
Y_{l'l}
\left[
\tilde{H}^{T}
(L_{lL})^{c}
\right]
+
\mathrm{h.c.}
,
\end{equation}
for $l,l'=e,\mu,\tau$.
Here
\begin{equation}\label{heavyH1}
L_{lL}
=
\left(
\begin{array}{c}
\nu_{lL}
\\
l_L
\end{array}
\right)
,
\qquad
H
=
\left(
\begin{array}{c}
H^{(+)}
\\
H^{(0)}
\end{array}
\right)
\end{equation}
are the lepton and Higgs doublets, $\tilde{H}=i\tau_{2}H^{*}$ is the conjugated
Higgs doublet, $(L_{lL})^{c} = C(\overline L_{lL})^{T}$ is the
(right-handed) charge-conjugated lepton doublet and $Y_{l'l}=Y_{ll'}$
are dimensionless constants (presumably of order one).
Here $C$ is the charge-conjugation matrix
(which satisfies the relations $C\gamma^{T}_{\alpha}C^{-1}=-\gamma_{\alpha}$ and $C^{T}=-C$).

The Lagrangian (\ref{effL1}) does not conserve the total lepton number $L$.
Let us stress that this is the only Lagrangian term with a dimension-five operator which can be built with the SM fields.

The electroweak symmetry is spontaneously broken by
the vacuum expectation value of the Higgs field
\begin{equation}\label{heavyH2}
\tilde{H}_{0}
=
\frac{1}{\sqrt{2}}
\left(
\begin{array}{c}
v
\\
0
\end{array}
\right)
.
\end{equation}
From Eqs.~(\ref{effL1}) and (\ref{heavyH2}),
we obtain the {\em left-handed Majorana neutrino mass term}
\begin{equation}\label{Mjmass}
\mathcal{L}^{\mathrm{M}}=-
\frac{1}{2}
\sum_{l',l}
\overline \nu_{l'L}
\,
M^{L}_{l'l}
\,
(\nu_{lL})^{c}
+
\mathrm{h.c.},
\end{equation}
where
\begin{equation}
M^{L}_{l'l}
=
\frac{v^2}{\Lambda}
\,
Y_{l'l}
\end{equation}
After the diagonalization of the symmetric matrix $Y$
through the transformation
\begin{equation}
Y = U \, y \, U^{T},
\quad
U^{\dag} U = 1,
\quad
y_{ik} = y_{i} \delta_{ik}
,
\end{equation}
we obtain
\begin{equation}\label{Mjmass1}
\mathcal{L}^{\mathrm{M}}
=
-
\frac{1}{2}
\sum_{i}m_{i} \bar\nu_{i} \nu_{i}.
\end{equation}
Here
\begin{equation}\label{Mjmass2}
m_{i}=\frac{v^2}{\Lambda} \, y_{i}
,
\end{equation}
and
\begin{equation}\label{Mjmass3}
\nu_{i}=\sum_{l}U_{il}^{\dag}\nu_{lL}+\sum_{l}(U_{il}^{\dag}\nu_{lL})^{c}.
\end{equation}
From Eq.~(\ref{Mjmass3}) it follows that the field $\nu_{i}$ satisfies
the Majorana condition
\begin{equation}\label{Mjmass4}
\nu_{i}=\nu_{i}^{c}=C\bar\nu_{i}^{T}.
\end{equation}
Thus,
{\em $\nu_{i}$ is the field of the Majorana neutrino with mass $m_{i}$} given by Eq.~(\ref{Mjmass2}).

From Eq.~(\ref{Mjmass3}), one can see that the flavor field $\nu_{lL}$ is
connected to $\nu_{iL}$ by the standard mixing relation
\begin{equation}\label{Majmass1}
\nu_{lL} = \sum_{i} U_{li} \, \nu_{iL},
\end{equation}
where $U$ is the unitary PMNS mixing matrix
given in Eq.~(\ref{MixMat}) in the standard parameterization,
including the diagonal matrix of Majorana phases.

The values of the neutrino masses are determined by the seesaw factor
$v^2/\Lambda$. Assuming that $m_{3}\simeq 5\cdot 10^{-2}$
eV (which is the largest neutrino mass in the case of a neutrino mass
hierarchy $m_{1}\ll m_{2}\ll m_{3}$), we have $\Lambda \simeq 10^{15}$
GeV. Thus, the standard seesaw mechanism of neutrino mass
generation explains the smallness of neutrino masses by a violation
of the total lepton number $L$ in interactions due to physics beyond the SM at a very large (GUT) scale.

The local effective Lagrangian (\ref{effL1}) can be obtained
by considering the possible existence of
heavy Majorana leptons $N_{i}$ with masses $M_{i} \gg v$,
which are
singlets of the $SU(2)_{L} \times U(1)_{Y}$ gauge group of the SM.
These heavy Majorana leptons can have the
lepton number-violating Yukawa interaction with the standard lepton and Higgs doublets
\begin{equation}\label{heavyH}
\mathcal{L}^{Y}_{I}
=
-
\sqrt{2}
\sum_{i,l}
Y_{li}\overline L_{lL} N_{iR} \tilde{H }+
\mathrm{h.c.}
\end{equation}
At electroweak energies,
the interaction (\ref{heavyH})
generates
the effective Lagrangian (\ref{effL1})
at second order of perturbation theory. We have
\begin{equation}\label{HeavyN}
\sum_{i}
Y_{l'i}\frac{1}{M_{i}}Y_{li}
=
\frac{1}{\Lambda}Y_{l'l}.
\end{equation}
From this relation it follows that the masses $M_{i}$ determine
the scale of new physics.

The seesaw mechanism based on the Lagrangian (\ref{heavyH}) is called
``type I seesaw''.
There are two other well-studied\cite{hep-ph/9805219} ways to generate
the effective Lagrangian $\mathcal{L}^{\rm{eff}}_{I}$
and, consequently, the left-handed Majorana mass term (\ref{Mjmass}):
through the interaction of
the lepton and Higgs doublets with a heavy triplet scalar boson
(type II seesaw) or with a heavy
Majorana
triplet fermion (type III seesaw).

Summarizing, if small neutrino masses are generated by
the standard seesaw mechanism, we have the following consequences:
\begin{enumerate}
\item Neutrinos with definite masses are truly neutral Majorana particles.
\item Neutrino masses are given by the seesaw relation (\ref{Mjmass2}).
Hence, the neutrino masses are suppressed with respect to the masses
of charged leptons and quarks,
which are proportional to $v$,
by the small ratio $v/\Lambda$.
\item The Majorana neutrino mass term is the only implication
at the electroweak scale of
a possible existence of heavy Majorana particles.
\item $CP$-violating decays of heavy Majorana particles in the early
Universe could be the origin of the baryon asymmetry of the Universe
(see Ref.\cite{0802.2962}).

\end{enumerate}

\section{On the theory of $0\nu\beta\beta$-decay}
\label{theory}

In this Section we present a
brief derivation of the matrix element of the
neutrinoless double-beta decay process in Eq.~(\ref{betabeta}),
assuming that this process is induced by Majorana neutrino masses
and mixing (see Refs.\cite{Doi:1985dx,Bilenky:1987ty,1001.1946}).

The standard effective Hamiltonian of the process has the form
\begin{equation}\label{effHam}
{\mathcal{H}}_{I}(x)= \frac{G_F}{\sqrt{2}} \, 2 \, \bar{e}_{L}(x)
\gamma_{\alpha} \nu_{eL}(x)
\,
j^{\alpha}(x) + \mathrm{h.c.}
\end{equation}
Here $G_F$ is the Fermi constant and $j^{\alpha}(x)$ is the hadronic
charged current which does not change strangeness. In terms of the quark
fields, the current $j^{\alpha}(x)$ has the form
\begin{equation}\label{effHam1}
j^{\alpha}(x)=2\cos\vartheta_{C}\bar u_{L}(x)\gamma^{\alpha} d_{L}(x).
\end{equation}
The mixed flavor field $\nu_{eL}(x)$ is given by the relation
(\ref{Majmass1}) with $l=e$:
\begin{equation}\label{emixfield}
\nu_{eL}(x)=\sum_{i}U_{ei} \, \nu_{iL}(x),
\end{equation}
where $U$ is the PMNS mixing matrix and
$\nu_{i}(x)$ is the field of the Majorana neutrino with mass $m_{i}$,
which satisfies the Majorana condition
(\ref{Mjmass4}).

The process (\ref{betabeta}) is of second order in $G_{F}$, with
the exchange of virtual neutrinos. The matrix element of the process is given by
\begin{eqnarray}\label{Smatelem}
&&
\langle f|S^2|i\rangle=-4 \left(\frac{G_F}{\sqrt{2}}\right)^2N_{p_1}N_{p_2}
\int d^{4}x_{1}d^{4}x_{2}
\sum_{i}
\bar u_{L}(p_1) e^{ip_{1}x_{1}} \gamma_{\alpha} U_{ei}
\nonumber\\
&&
\times
\langle 0|T(\nu_{iL}(x_{1}) \nu^{T}_{iL}(x_{2})|0\rangle
\gamma^{T}_{\beta} U_{ei}
\bar u^{T}_{L}(p_2)
e^{ip_{2}x_{2}}
\langle N_{f}|T(J^{\alpha}(x_{1})J^{\beta}(x_{2}))|N_{i} \rangle.
\end{eqnarray}
Here $p_{1}$ and $p_{2}$ are electron four-momenta, $J^{\alpha}(x)$ is the
hadronic charged current in the Heisenberg representation\footnote{In
Eq.~(\ref{Smatelem}) strong interactions are taken into account.},
$N_{i}$ and $N_{f}$ are the states of the initial and final
nuclei with respective four-momenta
$P_{i}=(E_{i}, \vec{p}_{i})$
and
$P_{f}=(E_{f}, \vec{p}_{f})$, and
$N_{p}^{-1}=(2\pi)^{3/2}\sqrt{2p^{0}}$ is the standard
normalization factor.

\begin{figure}[t!]
\begin{center}
\includegraphics*[width=0.25\linewidth]{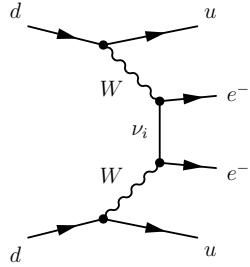}
\end{center}
\caption{\label{feybb1}
Feynman diagram of the elementary particle transition which induces
$0\nu\beta\beta$-decay.
}
\end{figure}

Taking into account the Majorana condition (\ref{Mjmass4}), for the
neutrino propagator we find the expression\footnote{The
neutrino propagator is proportional to $m_{i}$. This is connected to
the fact that only left-handed neutrino fields enter into the
Hamiltonian of weak interactions. Thus, in the case of massless
neutrinos the matrix element of neutrinoless double
$\beta$-decay is equal to zero. This is a consequence of the general
theorem on the equivalence of the theories with massless Majorana and
Dirac neutrinos\cite{Ryan-Okubo-NCS-2-234-1964,Case:1957}.}
\begin{equation}\label{nupropag}
\langle
0|T(\nu_{iL}(x_{1})
\bar\nu_{iL}(x_{2}))|0\rangle=-\frac{i}{(2\pi)^{4}}
\int d^{4}q \, e^{-iq(x_{1}-x_{2})}\frac{m_{i}}{q^2-m^2_{i}}
\,
\frac{1-\gamma_{5}}{2} \, C.
\end{equation}
Performing the integration over
$x^{0}_{1}$, $x^{0}_{2}$ and $q^{0}$ in Eqs.~(\ref{Smatelem}) and (\ref{nupropag}),
the matrix
element of the process takes the form
\begin{eqnarray}\label{Smatelem2}
&&\langle f|S^2|i
\rangle=2i \left(\frac{G_F}{\sqrt{2}}\right)^2N_{p_1}N_{p_2}
2\pi\delta(E_{f}+p^{0}_{1}+p^{0}_{2}-E_{i}) \bar
u(p_1)\gamma_{\alpha}\gamma_{\beta}(1+\gamma_{5})C \bar
u^{T}(p_2)\nonumber\\
&&
\times\int
d^{3}x_{1}
d^{3}x_{2}
e^{-i\vec{p}_{1}\vec{x}_{1}-i\vec{p}_{2}\vec{x}_{2}}
\sum_{j}
U^2_{ej}
m_{j}
\int \frac{d^{3}q}{(2\pi)^{3}} \,
\frac{e^{i\vec{q}(\vec{x}_{1}-\vec{x}_{2})}} {
q_{j}^{0}}\times\nonumber\\
&&\left(\sum_{n} \frac{\langle
N_{f}|J^{\alpha}(\vec{x}_{1})|N_{n}\rangle\langle N_{n}|
J^{\beta}(\vec{x}_{2}))|N_{i}\rangle }{E_{n}+p^{0}_{2}+q^{0}_{j}-E_{i}-i\epsilon}
+\sum_{n}\frac{ \langle
N_{f}|J^{\beta}(\vec{x}_{2})|N_{n}\rangle\langle N_{n}|
J^{\alpha}(\vec{x}_{1}))|N_{i}\rangle}
{E_{n}+p^{0}_{1}+q^{0}_{j}-E_{i}-i\epsilon} \right).
\nonumber\\
\end{eqnarray}
where
$
q^{0}_{j}
=
\sqrt{|\vec{q}|^2 + m_{j}^2}
$
and
$E_{n}$ are the energy levels of the intermediate nuclear state.

This is an exact expression for the matrix
element of $0\nu\beta\beta$-decay at second order of perturbation theory.
In the following we consider major $0^{+}\to 0^{+}$
transitions of even-even nuclei, for which
the following
standard approximations\cite{Doi:1985dx}
apply:

\begin{enumerate}

\item Effective Majorana mass approximation.

$0\nu\beta\beta$-decay is due to the exchange of virtual neutrinos (see the diagram
in Fig.\ref{feybb1}). Taking into account that the average distance between
nucleons in a nucleus is about $10^{-13}$ cm, the uncertainty
relation implies that the average neutrino momentum is $q \simeq 100$
MeV.
On the other hand, from tritium experiments we have the upper bounds in Eq.~(\ref{MainzTr}),
which constrain all the masses $m_{j}$ to be smaller than about 2 eV.
Therefore, the neutrino masses can be safely
neglected in the denominators in Eq.~(\ref{Smatelem2}) and we have
$q^{0}_{j} = \sqrt{|\vec{q}|^2 + m_{j}^2} \simeq q$,
with $q = |\vec{q}|$.

Thus, from Eq.~(\ref{Smatelem2}) it follows that in the matrix element of
$0\nu\beta\beta$-decay {\em the neutrino properties and the nuclear
properties are factorized and the neutrino masses and mixing enter into
the matrix element in the form of the effective Majorana mass}
\begin{equation}\label{effMj}
m_{\beta\beta}=\sum_{i}U^2_{ei}m_{i}.
\end{equation}

\item Long-wavelength approximation.

We have $|\vec{p}_{k}\vec{x}_{k}| \leq |\vec{p}_{k}| R$ ($k=1,2$), where
$R\simeq 1.2 \, A^{1/3}\cdot10^{-13}$ cm is the radius of a nucleus with nucleon number $A$.
Taking into account that $|\vec{p}_{k}| \lesssim 1 \, \mathrm{MeV}$, we have
$|\vec{p}_{k}\vec{x}_{k}| \ll 1$. Thus, we have
$e^{-i\vec{p}_{1}\vec{x}_{1}-i\vec{p}_{2} \vec{x}_{2}}\simeq 1$ (this
approximation means that electrons are produced in $S$-states).

\item Closure approximation.

The energy of the virtual neutrino, $q\simeq 100$ MeV, is much larger than
the excitation energy $E_{n}-E_{i}$. Thus, the energy of the
intermediate states $E_{n}$ can be approximated by an average energy
$\overline{E}$.
In this approximation,
called ``closure approximation'',
we have
\begin{equation}\label{Smatelem3}
\frac{\langle
N_{f}|J^{\alpha}(\vec{x}_{1})|N_{n}\rangle\langle N_{n}|
J^{\beta}(\vec{x}_{2}))|N_{i}\rangle }{E_{n}+p^{0}_{k}+q^{0}_{j}-E_{i}-i\epsilon}\simeq \frac{\langle
N_{f}|J^{\alpha}(\vec{x}_{1})
J^{\beta}(\vec{x}_{2}))|N_{i}\rangle }{\overline{ E}+p^{0}_{k}+q-E_{i}-i\epsilon}.
\end{equation}

\end{enumerate}

Taking into account these approximations
and considering commuting hadronic currents
(see Eqs.~(\ref{current1}) and (\ref{current2}) below),
for the matrix element of
$0\nu\beta\beta$-decay we obtain the expression
\begin{eqnarray}\label{Smat1}
&&\langle f|S^{(2)}|i
\rangle=8\pi i \left(\frac{G_F}{\sqrt{2}}\right)^2 m_{\beta\beta}N_{p_1}N_{p_2}
\bar
u(p_1)(1+\gamma_{5})C \bar
u^{T}(p_2)\times \nonumber\\
&&\int d^{3}x_{1}d^{3}x_{2} \langle
N_{f}|J^{\alpha}(\vec{x}_{1})K(|\vec{x}_{1}-\vec{x}_{2}|)
J_{\alpha}(\vec{x}_{2}))|N_{i}\rangle
\delta(E_{f}+p^{0}_{1}+p^{0}_{2}-E_{i}).\nonumber\\
\end{eqnarray}
Here
\begin{equation}\label{Smat2}
K(|\vec{x}_{1}-\vec{x}_{2}|)=\frac{1}{(2\pi)^{3}}
\int
d^{3}q
\,
\frac
{e^{i\vec{q} (\vec{x}_{1}-\vec{x}_{2})}}
{q \Big[ \overline{E} + q - \left(M_{i}+M_{f}\right)/2 \Big]}
,
\end{equation}
where $M_{i}(M_{f})$ is the mass of the initial (final) nucleus.

In the calculation of the hadronic part of the matrix element of
$0\nu\beta\beta$-decay, the following approximate expression
for the effective charged current
$J^{\alpha}(\vec{x})=(J^{0}(\vec{x}),\vec{ J}(\vec{x}))$
is used\cite{hep-ph/9905509}:
\begin{equation}\label{current1}
J^{0}(\vec{x}) = \sum^{A}_{n=1}\tau_{n}^{+}\delta(\vec{x}-\vec{r}_{n})
g_{V}(q^2)
\end{equation}
and
\begin{equation}\label{current2}
\vec{ J}(\vec{x}) =
-\sum^{A}_{n=1}\tau_{n}^{+}\delta(\vec{x}-\vec{r}_{n})
\left[
g_{A}(q^2)\vec{\sigma}_{n}+g_{M}(q^2)i\frac{\vec{\sigma}_{n}
\times \vec{q}}{2m_{p}}-g_{P}(q^2)
\frac{\left(\vec{\sigma}_{n}\cdot\vec{q}\right) \vec{q}}{2m_{p}}
\right].
\end{equation}
Here, $\sigma_{n}^{i}$ and $\tau_{n}^{i}$ are Pauli matrices
acting, respectively, on the spin and isospin doublets of the $n$ nucleon,
$\tau_{+}=(\tau_{1}+i \tau_{2})/2$, $\vec{r}_{n}$ is the
coordinate of the $n$ nucleon,
$m_{p}$ is the proton mass,
$g_{V}(q^2), g_{A}(q^2), g_{M}(q^2)$ and $g_{P}(q^2)$
are the vector, axial,
magnetic and pseudoscalar weak form factors of the nucleon. From the
conserved vector current (CVC) and partially conserved axial current
(PCAC) hypotheses, it follows that
\begin{equation}\label{current3}
g_{V}(q^2)=F^{p}_{1}(q^2)-F^{n}_{1}(q^2),
\quad
g_{M}(q^2)=F^{p}_{2}(q^2)-F^{n}_{2}(q^2),
\quad
g_{P}(q^2)=\frac{2m_{p}g_{A}}{q^2+m^2_{\pi}},
\end{equation}
where $F^{p(n)}_{1}$ and $F^{p(n)}_{2}$ are the Dirac and Pauli
electromagnetic form factors
of the proton (neutron) and $g_{A}\simeq 1.27$ is the axial coupling
constant of the nucleon.

The expressions (\ref{current1}) and (\ref{current2})
can be obtained from the one-nucleon matrix element
of the hadronic charged current. For the number density of nucleons
in a nucleus, the following approximate expression is used:
\begin{equation}\label{current4}
\bar\Psi(\vec{x})\gamma^{0}\Psi(\vec{x})=
\sum^{A}_{n=1}\delta(\vec{x}-\vec{r}_{n}).
\end{equation}

The nuclear matrix element (NME)
$M^{0\nu}$, which is the integrated product of two hadronic charged currents
and a neutrino propagator,
is a sum of a Fermi (F), a Gamow-Teller (GT) and a tensor (T)
term:
\begin{equation}\label{MatEl}
{M}^{0\nu} = \langle 0^+_f|\sum_{k,l} \tau^+_k \tau^+_l \left[
\frac{H_F(r_{kl})}{g^2_A} + H_{GT}(r_{kl}) \vec\sigma_k\cdot
\vec\sigma_l - H_T(r_{kl}) S_{kl} \right]
|0^+_i\rangle
.
\end{equation}
Here
$S_{kl} = 3(\vec\sigma_k\cdot \vec{r}_{kl})
(\vec\sigma_l \cdot \vec{r}_{kl})
- \vec\sigma_k\cdot \vec\sigma_l$,
with
$\vec{r}_{kl}=\vec{r}_{k}-\vec{r}_{l}$,
and the neutrino potentials $H_{F,GT,T}(r_{kl})$ are
given by the expressions
\begin{equation}
H_{F,GT,T}(r_{kl}) = \frac{2}{\pi} R
\int_0^\infty \frac{j_{0,0,2}(q r_{kl}) h_{F,GT,T}(q^2) q}{q + \overline{E} -(M_i+M_f)/2 }
dq,
\end{equation}
where $R$ is the radius of the nucleus, and the functions
$h_{F,GT,T}(q^2)$ are combinations of different form factors\footnote{The functions $h_{F,GT,T}(q^2)$ can be found in
Ref.\cite{0710.2055}.}.

Taking into account the Coulomb interaction of the electrons and the final
nucleus, for the total width of $0\nu\beta\beta$-decay we find the general expression
\begin{equation}\label{totrate}
\Gamma^{0\nu}
=
\frac{1}{T^{0\nu}_{1/2}}
=
G^{0\nu}(Q,Z)
\,
|M^{0\nu}|^2
\,
\frac{|m_{\beta\beta}|^2}{m_{e}^2}
,
\end{equation}
where $G^{0\nu}(Q,Z)$ is a known integral over the phase space,
$
Q
=
M_{i}
-
M_{f}
-
2 \, m_{e}
$
is the $Q$-value of the process,
and
$m_{e}$ is the electron mass.
The numerical values of $G^{0\nu}(Q,Z)$, $Q$ and the natural abundance of
several nuclei of experimental interest are
presented in Table~\ref{tabnuc}.

\begin{table}[b!]
\tbl{The values of $G^{0\nu}(Q,Z)$, $Q$ and natural abundance of the initial isotope for
several $\beta\beta$-decay processes of experimental interest.
Table adapted from Ref.\protect\cite{Schwingenheuer:2012jt}.}
{
\begin{tabular}{ccccl}
$\beta\beta$-decay & $G^{0\nu}$ & $Q$ & nat.~abund. & experiments \\
 & $[10^{-14}\,\mathrm{y}^{-1}]$ & [keV] & [\%] & \\ \hline
$ {}^{48}\text{Ca} \to {}^{48}\text{Ti} $ & 6.3 & 4273.7 & 0.187 & CANDLES \\
$ {}^{76}\text{Ge} \to {}^{76}\text{Se} $ & 0.63 & 2039.1 & 7.8 & GERDA, Majorana \\
$ {}^{82}\text{Se} \to {}^{82}\text{Kr} $ & 2.7 & 2995.5 & 9.2 & SuperNEMO, Lucifer \\
$ {}^{100}\text{Mo} \to {}^{100}\text{Ru} $ & 4.4 & 3035.0 & 9.6 & MOON, AMoRe \\
$ {}^{116}\text{Cd} \to {}^{116}\text{Sn} $ & 4.6 & 2809 & 7.6 & Cobra \\
$ {}^{130}\text{Te} \to {}^{130}\text{Xe} $ & 4.1 & 2530.3 & 34.5 & CUORE \\
$ {}^{136}\text{Xe} \to {}^{136}\text{Ba} $ & 4.3 & 2461.9 & 8.9 & EXO, KamLAND-Zen, NEXT, XMASS \\
$ {}^{150}\text{Nd} \to {}^{150}\text{Sm} $ & 19.2 & 3367.3 & 5.6 & SNO+, DCBA/MTD \\
\end{tabular}
\label{tabnuc}
}
\end{table}


\section{Effective Majorana mass}
\label{effmass}

The effective Majorana mass $m_{\beta\beta}$ is determined by
the neutrino masses, the mixing angles and the Majorana phases.
In this Section
we discuss which are the possible values of the effective Majorana mass
which can be obtained taking into account the information on the neutrino mass-squared differences
and mixing angles obtained from neutrino oscillation data.

In the standard parameterization (\ref{MixMat}) of the mixing matrix, we have
\begin{equation}
|m_{\beta\beta}|
=
\left|
\cos^2\vartheta_{12} \cos^2\vartheta_{13} m_{1}
+
e^{2i\alpha_{12}}
\sin^2\vartheta_{12} \cos^2\vartheta_{13} m_{2}
+
e^{2i\alpha_{13}}
\sin^2\vartheta_{13} m_{3}
\right|
\,,
\label{mbbst}
\end{equation}
where $\alpha_{12}$ and $\alpha_{13}$
are, respectively, the phase differences of $U_{e2}$ and $U_{e3}$
with respect to $U_{e1}$:
$\alpha_{12}=\lambda_{2}$
and
$\alpha_{13}=\lambda_{3}-\delta$
in the standard parameterization (\ref{MixMat}) of the mixing matrix.
Therefore,
$0\nu\beta\beta$-decay depends not only on the mixing angles and Dirac CP-violating phase,
but also on the Majorana CP-violating phases.
This is in agreement with the discussion after Eq.~(\ref{MixMat}),
since the total lepton number is violated in $0\nu\beta\beta$-decay.

\begin{figure}[t!]
\begin{minipage}[r]{0.47\textwidth}
\begin{center}
Before Daya Bay
\\
\includegraphics*[width=0.99\textwidth]{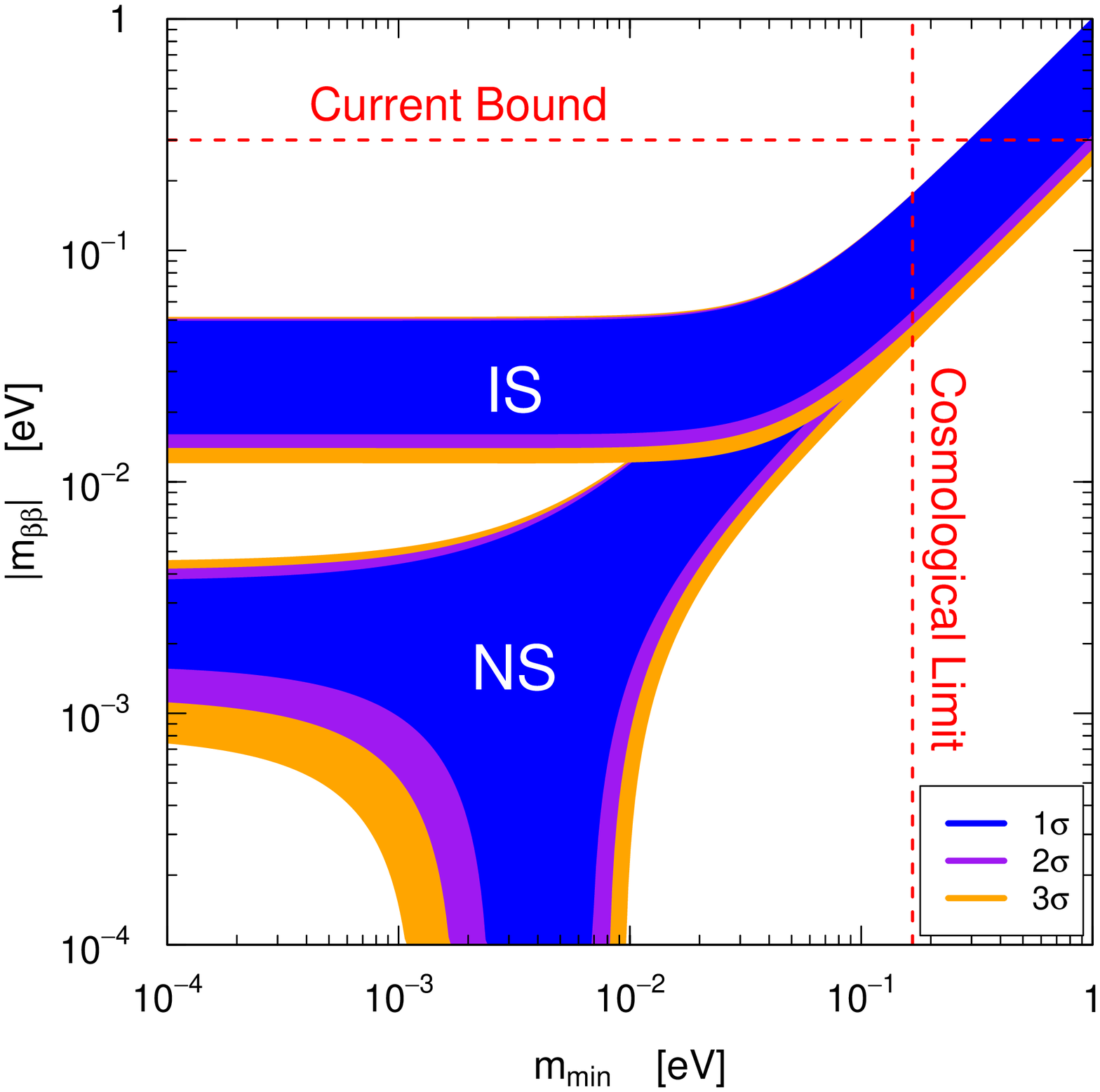}
\end{center}
\end{minipage}
\hfill
\begin{minipage}[l]{0.47\textwidth}
\begin{center}
After Daya Bay
\\
\includegraphics*[width=0.99\textwidth]{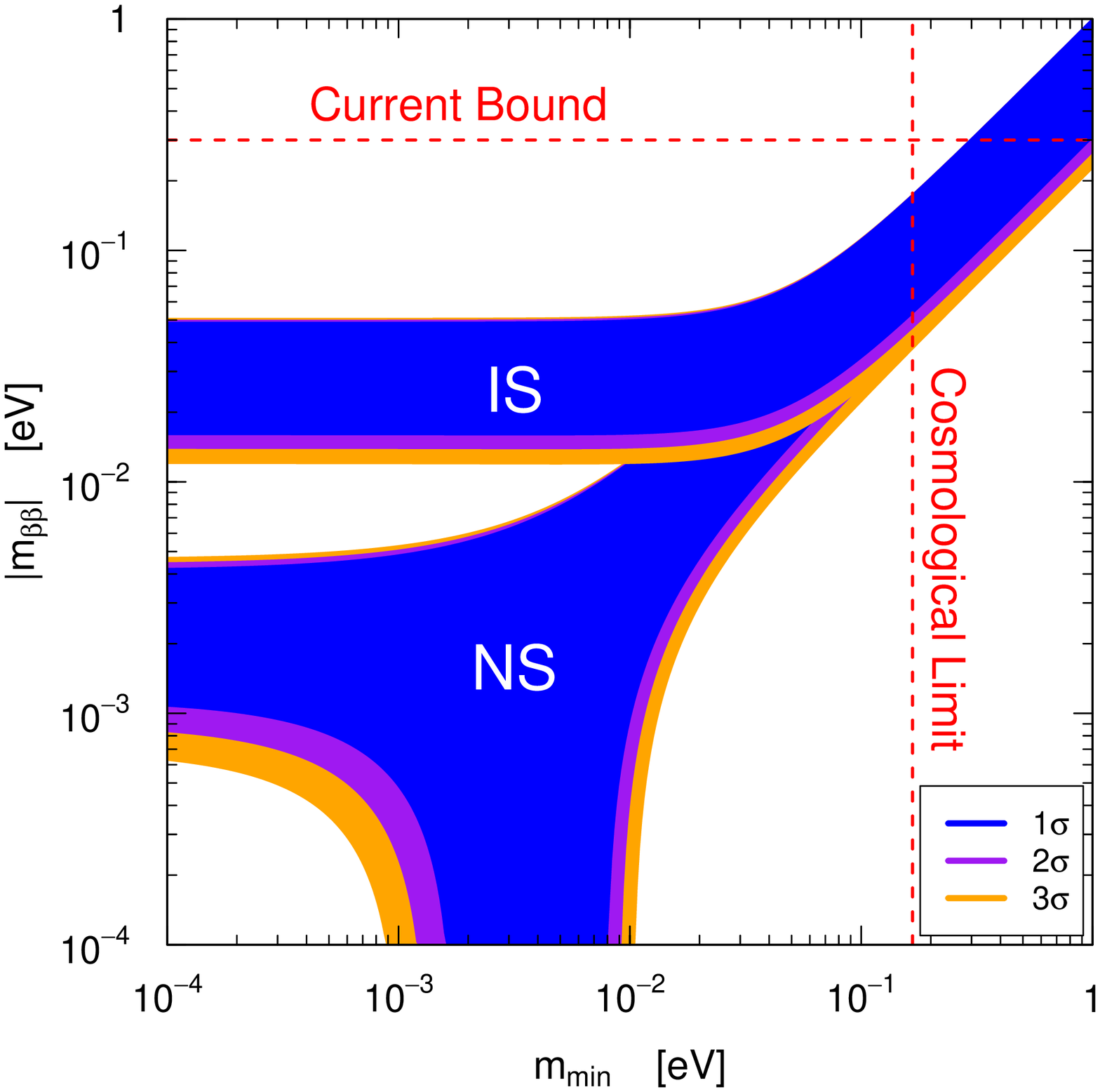}
\end{center}
\end{minipage}
\caption{\label{bb0-plt}
Value of the effective Majorana mass $|m_{\beta\beta}|$ as a
function of the lightest neutrino mass
in the normal (NS, with $m_{\mathrm{min}}=m_{1}$)
and
inverted (IS, with $m_{\mathrm{min}}=m_{3}$)
neutrino mass spectra
before and after the Daya Bay\protect\cite{1203.1669}
measurement of $\vartheta_{13}$ in Eq.~(\ref{s13db}).
The current upper bound on $|m_{\beta\beta}|$
(see Eqs.~(\ref{Hei-Mos1}), (\ref{Cuori1}) and (\ref{Nemo1}))
and 
the cosmological bound
(see Ref.\protect\cite{1007.0658})
on $\sum_{i} m_{i} \simeq 3 m_{\mathrm{min}}$
in the quasi-degenerate region
are indicated.
}
\end{figure}

In the case of a NS,
the neutrino masses $m_{2}$ and $m_{3}$
are connected with
the lightest mass $m_{1}$
by the relations
\begin{equation}\label{norspec1}
m_{2}=\sqrt{m^2_{1}+\Delta{m}^2_{s}},
\qquad
m_{3}=\sqrt{m^2_{1}+\Delta{m}^2_{s}+\Delta{m}^2_{a}}.
\end{equation}
On the other hand,
in a IS $m_{3}$ is the lightest mass and we have
\begin{equation}\label{invspec1}
m_{1}=\sqrt{m^2_{3}+\Delta{m}^2_{a}},
\qquad
m_{2}=\sqrt{m^2_{3}+\Delta{m}^2_{a}+\Delta{m}^2_{s}}.
\end{equation}

Figure~\ref{bb0-plt} shows the value of the effective Majorana mass $|m_{\beta\beta}|$ as a
function of the lightest neutrino mass\cite{hep-ph/9906525,hep-ph/0102265}
in the normal and inverted neutrino mass spectra
before and after the Daya Bay\cite{1203.1669}
measurement of $\vartheta_{13}$ in Eq.~(\ref{s13db}).
We used the values of the
neutrino oscillation parameters obtained in the global analysis presented in Ref.\cite{Schwetz:2011zk}:
\begin{equation}
\Delta{m}^2_{12}
=
7.59 {}^{+(0.20,0.40,0.60)}_{-(0.18,0.35,0.50)} \times 10^{-5} \, \text{eV}^2
,
\;
\sin^2 \vartheta_{12}
=
0.312 {}^{+(0.017,0.038,0.058)}_{-(0.015,0.032,0.042)}
,
\label{arXiv:1103.0734-1}
\end{equation}
and in the NS
\begin{equation}
\Delta{m}^2_{13}
=
2.50 {}^{+(0.09,0.18,0.26)}_{-(0.16,0.25,0.36)} \times 10^{-3} \, \text{eV}^2
,
\;
\sin^2 \vartheta_{13}
=
0.013 {}^{+(0.007,0.015,0.022)}_{-(0.005,0.009,0.012)}
,
\label{arXiv:1103.0734-2}
\end{equation}
whereas in the IS
\begin{equation}
-
\Delta{m}^2_{13}
=
2.40 {}^{+(0.08,0.18,0.27)}_{-(0.09,0.17,0.27)} \times 10^{-3} \, \text{eV}^2
,
\;
\sin^2 \vartheta_{13}
=
0.016 {}^{+(0.008,0.015,0.023)}_{-(0.006,0.011,0.015)}
.
\label{arXiv:1103.0734-3}
\end{equation}
The three levels of uncertainties correspond to $(1\sigma,2\sigma,3\sigma)$.
In the ``After Daya Bay'' plot in Fig~\ref{bb0-plt}
we replaced the value of $\vartheta_{13}$ in Eqs.~(\ref{arXiv:1103.0734-2}) and (\ref{arXiv:1103.0734-3})
with that measured by the Daya Bay Collaboration in Eq.~(\ref{s13db}).
The uncertainties for
$|m_{\beta\beta}|$ have been calculated using the standard method of propagation of uncorrelated errors,
taking into account the asymmetric uncertainties in Eqs.~(\ref{arXiv:1103.0734-1})--(\ref{arXiv:1103.0734-3}).

In the following we discuss the predictions for the effective Majorana mass in three cases with characteristic neutrino mass spectra:
\begin{enumerate}
\item Hierarchy of neutrino masses\footnote{Quarks and charged leptons have this type of mass spectrum.}:
\begin{equation}\label{hierar}
m_{1} \ll m_{2} \ll m_{3}.
\end{equation}
\item Inverted hierarchy of neutrino masses:
\begin{equation}\label{invierar}
m_{3} \ll m_{1} \lesssim m_{2}.
\end{equation}
\item
Quasi-degenerate neutrino mass spectrum:
\begin{equation}\label{quasi}
\sqrt{ \Delta{m}^2_{\rm{a}}}
\ll
m_{0}
\simeq
\left\{
\begin{array}{rcl} \displaystyle
m_{1}
\lesssim
m_{2}
\lesssim
m_{3}
& \displaystyle
\qquad
& \displaystyle
\mathrm{(NS)},
\\ \displaystyle
m_{3}
\lesssim
m_{1}
\lesssim
m_{2}
& \displaystyle
\qquad
& \displaystyle
\mathrm{(IS)},
\end{array}
\right.
\end{equation}
\end{enumerate}
where $m_{0}$ is the absolute mass scale common to the three masses.
As one can see from Fig.~\ref{bb0-plt},
the Daya Bay measurement of $\vartheta_{13}$
has a visible impact on the value of $|m_{\beta\beta}|$
only in the case of a hierarchy of neutrino masses,
discussed in the following,
because only in that case the contribution of the largest mass $m_{3}$,
which is weighted by $\sin^2\vartheta_{13}$,
is decisive.

\subsection{Hierarchy of neutrino masses}
\label{hierarchy}

In this case we have
\begin{equation}\label{hierar1}
m_{1} \ll \sqrt{\Delta{m}^2_{s}},
\qquad
m_{2}\simeq \sqrt{ \Delta{m}^2_{s}},
\qquad
m_{3}\simeq \sqrt{ \Delta{m}^2_{a}}.
\end{equation}
Thus,
$m_{2}$ and $m_{3}$ are determined by the solar and atmospheric
neutrino mass-squared differences.
Neglecting the
contribution of $m_{1}$ to the effective Majorana mass,
from Eq.~(\ref{mbbst})
we find
\begin{equation}\label{hierar2}
|m_{\beta\beta}|
\simeq
\left|
\sin^2\vartheta_{12} \sqrt{\Delta{m}^2_{s}}
+
e^{2i\alpha_{23}}
\sin^2\vartheta_{13} \sqrt{\Delta{m}^2_{a}}
\right|
,
\end{equation}
where $\alpha_{23}$ is the phase difference between $U_{e3}$ and $U_{e2}$:
$\alpha_{23}=\alpha_{13}-\alpha_{12}=\lambda_{3}-\delta-\lambda_{2}$
in the standard parameterization (\ref{MixMat}) of the mixing matrix.

The first term in Eq.(\ref{hierar2}) is small because of the
smallness of $\Delta{m}^2_{s}$.
On the other hand,
the contribution of the ``large''
$\Delta{m}^2_{a}$ is suppressed by the small factor $\sin^2
\vartheta_{13} $.
Hence,
both terms must be taken into account and cancellations are possible,
as shown in Fig.~\ref{bb0-plt}.

As one can see from Fig.~\ref{bb0-plt},
in the case of a hierarchy of neutrino masses we have
the upper bound
\begin{equation}\label{hierar4}
|m_{\beta\beta}|
\leq
\sin^2\vartheta_{12} \sqrt{\Delta{m}^2_{s}}
+
\sin^2\vartheta_{13} \sqrt{\Delta{m}^2_{a}}
\lesssim
5 \cdot 10^{-3} \, \mathrm{eV},
\end{equation}
which is significantly smaller than the expected
sensitivity of the future experiments on the search for
$0\nu\beta\beta$-decay (see Section~\ref{exp}).
This bound corresponds to the case of
$e^{2i\alpha_{23}}=1$.
It is slightly increased by the Daya Bay
measurement of $\vartheta_{13}$ in Eq.~(\ref{s13db}),
because the additive contribution of $\sin^2\vartheta_{13} \sqrt{\Delta{m}^2_{a}}$ in Eq.~(\ref{hierar2}) is increased.
On the other hand,
one can see from Fig.~\ref{bb0-plt} that
the lower bound on $|m_{\beta\beta}|$ for $m_{1} \ll 10^{-3} \, \mathrm{eV}$,
which corresponds to
$e^{2i\alpha_{23}}=-1$,
is slightly decreased by the Daya Bay
measurement of $\vartheta_{13}$,
because the increased contribution of $\sin^2\vartheta_{13} \sqrt{\Delta{m}^2_{a}}$ in this case is subtracted.

From Fig.~\ref{bb0-plt} one can also see that
when the contribution of $m_{1}$ is not negligible,
there can be cancellations among the three mass contributions.
The two extreme cases in which cancellations can happen are the following ones in which CP is conserved:

\begin{description}

\item[$e^{2i\alpha_{12}}=-1\;\mathrm{and}\;e^{2i\alpha_{13}}=+1.$]
The value of $m_{1}$ for which cancellations suppress
$|m_{\beta\beta}|$
is slightly decreased by the Daya Bay
measurement of $\vartheta_{13}$, because the larger value of $\sin^2\vartheta_{13} m_{3}$
adds to the contribution of $m_{1}$.
Hence, a smaller value of $m_{1}$ is required to cancel the sum of the
contributions of $m_{1}$ and $m_{3}$
with the opposite contribution of $m_{2}$.

\item[$e^{2i\alpha_{12}}=-1\;\mathrm{and}\;e^{2i\alpha_{13}}=-1.$]
The value of $m_{1}$ for which cancellations suppress
$|m_{\beta\beta}|$
is slightly increased by the Daya Bay
measurement of $\vartheta_{13}$, because the larger value of $\sin^2\vartheta_{13} m_{3}$
adds to the contribution of $m_{2}$.
Hence, a larger value of $m_{1}$ is required to cancel the contribution of $m_{1}$
with the opposite sum of contributions of $m_{2}$ and $m_{3}$.

\end{description}

Figure~\ref{bb0-plt} shows that\footnote{
We are very grateful to Michele Frigerio for pointing out a mistake in the cancellation band
presented in the first arXiv version of this paper
and in its published version
(Mod. Phys. Lett. A 27 (2012) 1230015).
}
the two effects lead to a slight widening of the cancellation band after the Daya Bay measurement of $\vartheta_{13}$.

\subsection{Inverted hierarchy of the neutrino masses}
\label{inverted}

In this case,
for the neutrino masses we have
\begin{equation}\label{invierar1}
m_{3}
\ll
\sqrt{\Delta{m}^2_{a}},
\quad
m_{1}
\simeq
\sqrt{\Delta{m}^2_{a}},
\quad
m_{2}
\simeq
\sqrt{\Delta{m}^2_{a}}
\left(
1+
\frac{\Delta{m}^2_{s}}{2 \Delta{m}^2_{a}}
\right)
\simeq
\sqrt{ \Delta{m}^2_{a}}.
\end{equation}
In the expression of $|m_{\beta\beta}|$,
the contribution of the term $m_{3}\sin^2\vartheta_{13}$ can be safely
neglected. Neglecting also the small contribution of
$\sin^2\vartheta_{13}$,
from Eq.~(\ref{mbbst})
we find
\begin{equation}\label{invierar2}
|m_{\beta\beta}|
\simeq
\sqrt{\Delta{m}^2_{a} \left(1-\sin^22\vartheta_{12}\,\sin^2\alpha_{12}\right)}.
\end{equation}
The phase $\alpha_{12}$ is the only unknown parameter
in the expression for the effective Majorana mass in the case of
a inverted mass hierarchy.

From Eq.~(\ref{invierar2}) we find the following range for $|m_{\beta\beta}|$:
\begin{equation}\label{invierar3}
\cos 2\vartheta_{12} \,\sqrt{ \Delta{m}^2_{a}} \leq
|m_{\beta\beta}| \leq\sqrt{ \Delta{m}^2_{a}}.
\end{equation}
The upper and lower bounds of this inequality
correspond to the case of $CP$-invariance in the lepton sector.
In fact, $CP$ invariance implies that
(see Refs.\cite{Bilenky:1987ty,Giunti-Kim-2007,Bilenky:2010zza})
\begin{equation}\label{CPMaj1}
e^{2i\alpha_{12}}=\eta_{2} \, \eta^{*}_{1},
\end{equation}
where $\eta_{k}=\pm i$ is the $CP$ parity of the Majorana neutrino $\nu_{k}$.
If $\eta_{2}=\eta_{1}$, we have $\alpha_{12}=0,\pi$ (the upper bound
in the inequality (\ref{invierar3})).
If $\eta_{2}=-\eta_{1}$ we have
$\alpha_{12}=\pm\pi/2$ (the lower bound in the inequality
(\ref{invierar3})).

From the existing neutrino oscillation data, we find the following
range for the possible value of the effective Majorana mass:
\begin{equation}\label{invierar5}
10^{-2} \lesssim |m_{\beta\beta}| \lesssim 5 \cdot 10^{-2} \, \mathrm{eV}.
\end{equation}
The anticipated sensitivities to $|m_{\beta\beta}|$ of
the future experiments on the search
for the $0\nu\beta\beta$-decay are
in the range (\ref{invierar5}) (see Section~\ref{exp}).
Thus, the future
$0\nu\beta\beta$-decay experiments will probe the Majorana nature
of neutrinos if a inverted hierarchy of neutrino masses is realized in nature.

\subsection{Quasi-degenerate neutrino mass spectrum}
\label{quasi-degenerate}

Neglecting the small contribution of $\sin^2\vartheta_{13}$ in Eq.~(\ref{mbbst}),
in the case of a quasi-degenerate neutrino mass spectrum
we obtain
\begin{equation}\label{quasi1}
|m_{\beta\beta}|
\simeq
m_{0}
\sqrt{1-\sin^22\vartheta_{12}\,\sin^2\alpha_{12}},
\end{equation}
where $m_{0}$
is the unknown absolute mass scale of neutrino masses
(see Eq.~(\ref{quasi}))
and $\alpha_{12}$ is the phase difference between $U_{e2}$ and $U_{e1}$:
$\alpha_{12}=\lambda_{2}$
in the standard parameterization (\ref{MixMat}) of the mixing matrix.
Thus, in this case
$|m_{\beta\beta}|$ depends
on two
unknown parameters: $m_{0}$ and $\alpha_{12}$.

From Eq.~(\ref{quasi1}), we obtain the following range for
the effective Majorana mass:
\begin{equation}\label{quasi2}
\cos 2\vartheta_{12} \, m_{0}
\leq
|m_{\beta\beta}|
\leq
m_{0}
.
\end{equation}
If $0\nu\beta\beta$-decay will be observed and the effective Majorana
mass will turn out to be relatively large
($|m_{\beta\beta}| \gg \sqrt{\Delta{m}^2_{a}}$),
we will have an evidence that neutrinos are
Majorana particles and their mass spectrum is
quasi-degenerate. In this case, we have
\begin{equation}\label{quasi5}
|m_{\beta\beta}|
\leq
m_{0}
\leq
\frac{|m_{\beta\beta}|}{\cos 2\vartheta_{12}}
\simeq
2.8\, |m_{\beta\beta}|.
\end{equation}

Information about the value of the mass scale will be inferred from
the data of the future tritium $\beta$-decay experiment KATRIN
\cite{hep-ex/0109033,Angrik:2005ep} and from future cosmological observations.
The sensitivity
of the KATRIN experiment to the neutrino mass scale is
expected to be about 0.2 eV,
which the same as the sensitivity to
$m_{\beta}$
in Eq.~(\ref{trimass}),
since in the quasi-degenerate case
$m_{\beta} \simeq m_{0}$.
Cosmological observations give information on the value of the sum of the
neutrino masses $\sum_{i}m_{i}\simeq3m_{0}$
in the quasi-degenerate case.
The existing cosmological data imply the
bound $\sum_{i}m_{i}\lesssim 0.5$ eV
(see Ref.\cite{1007.0658}).
It is expected that future cosmological
observations will be sensitive to $\sum_{i}m_{i}$ in the range
$(6\times 10^{-3}-10^{-1})$ eV (see, for example, Ref.\cite{1103.5083}).

\section{Nuclear matrix elements}
\label{nucmatel}

\begin{figure}[t!]
\begin{center}
\includegraphics*[width=0.8\textwidth]{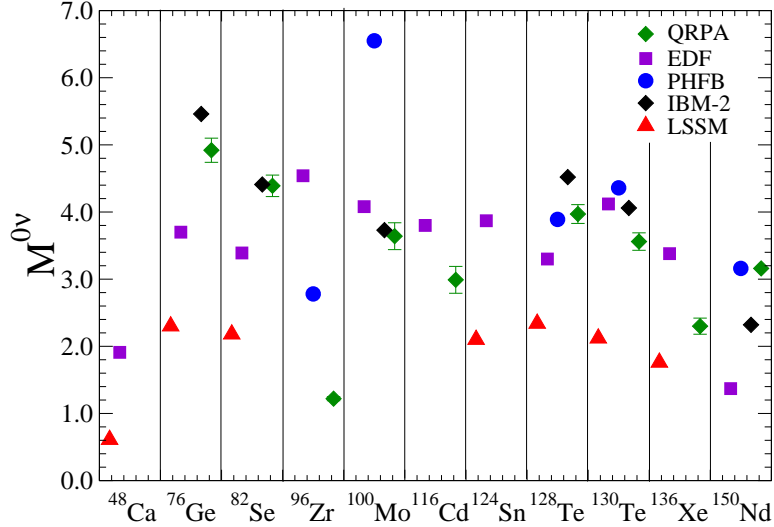}
\end{center}
\caption{\label{simkovic-nmes}
Values of the NME calculated with the methods in Tab.~\ref{tabnme}~\protect\cite{Simkovic-private-12}.
}
\end{figure}

The effective Majorana mass $|m_{\beta\beta}|$ is not a directly
measurable quantity.
The measurement of the half-life of $0\nu\beta\beta$-decay gives
{\em the product of the effective Majorana mass and the nuclear matrix element}
(see Eq.~(\ref{totrate})).
Hence,
in order to determine the effective Majorana mass
one must calculate the nuclear matrix elements (NMEs) of $0\nu\beta\beta$-decay,
which is a complicated nuclear many-body problem.
Five different methods are used at present.
In this short review we do not describe these methods and
we do not discuss the advantages and disadvantages
of each of them.
We only present the references to the original papers in Tab.~\ref{tabnme} and the latest results
in Fig.~\ref{simkovic-nmes}.

\begin{table}[b!]
\tbl{Methods of calculation of nuclear matrix elements of $0\nu\beta\beta$-decay.}
{
\begin{tabular}{cc}
Method & References
\\
\hline
Quasi-particle Random Phase Approximation (QRPA) & \cite{nucl-th/0305005,0706.4304,nucl-th/0012010,Simkovic:2011zz}
\\
Energy Density Functional method (EDF) & \cite{1012.1783,1008.5260}
\\
Projected Hartree-Fock-Bogoliubov approach (PHFB) & \cite{Rath:2010zz,Rath:2011zz}
\\
Interacting Boson Model-2 (IBM-2) & \cite{Barea:2009zz,Iachello:2011zz,Iachello:2011zzb}
\\
Large-Scale Shell Model (LSSM) & \cite{Menendez:2011zza,0801.3760}
\end{tabular}
\label{tabnme}
}
\end{table}

From Fig.~\ref{simkovic-nmes} we reach the following conclusions:

\begin{enumerate}

\item
The LSSM value of each NME is typically smaller than
the corresponding one calculated with other approaches.
Moreover, the
LSSM value of each NME
depends weakly on the nucleus,
except for the double-magic nucleus ${^{48}\rm{Ca}}$.
If $0\nu\beta\beta$-decay of
different nuclei will be observed in future experiments, this
characteristic feature of the LSSM can be checked,
because the LSSM predicts the following ratio of half-lives of different nuclei:
\begin{equation}
\frac{T^{0\nu}_{1/2}(Z_{1},A_{1})}{T^{0\nu}_{1/2}(Z_{2},A_{2})}
\simeq
\frac{G^{0\nu}(Q_{2},Z_{2})}{G^{0\nu}(Q_{1},Z_{1})}
\end{equation}

\item
\label{item:ratnme}
There is a large discrepancy between the values of NMEs
calculated with different approaches.
The ratios of
the maximal and minimal values of each NME are
3.1 ($^{48}\mathrm{Ca}$),
2.4 ($^{76}\mathrm{Ge}$),
2.0 ($^{82}\mathrm{Se}$),
3.7 ($^{96}\mathrm{Zr}$),
1.8 ($^{100}\mathrm{Mo}$),
1.3 ($^{116}\mathrm{Cd}$),
1.8 ($^{124}\mathrm{Sn}$),
1.9 ($^{128}\mathrm{Te}$),
2.1 ($^{130}\mathrm{Te}$),
1.9 ($^{136}\mathrm{Xe}$),
2.3 ($^{150}\mathrm{Nd}$).
Therefore,
the situation with the calculation of the
$0\nu\beta\beta$-decay NMEs is obviously not satisfactory at present.
Further efforts and progress are definitely needed.

\end{enumerate}

\section{Neutrinoless double-beta decay experiments}
\label{exp}

Many experiments searched for neutrinoless double-beta decay
without finding an uncontroversial positive evidence.
The most stringent lower bounds on the
half-lives of the decays of $^{76}\mathrm{Ge}$, $^{130}\mathrm{Te}$
and $^{100}\mathrm{Mo}$ have been obtained, correspondingly, in the
Heidelberg-Moscow\cite{Klapdor-Kleingrothaus:2001yx}, Cuoricino\cite{1012.3266} and
NEMO3\cite{hep-ex/0601021,Barabash:2010bd} experiments.

In the Heidelberg-Moscow experiment\cite{Klapdor-Kleingrothaus:2001yx}
Germanium crystals with a
86\% enrichment in the $\beta\beta$-decaying isotope $^{76}\mathrm{Ge}$ 
were used.
The total mass of $^{76}\mathrm{Ge}$ was 11 kg,
with a low background of 0.11 counts/(kg~keV~y).
After 13 years of running (with a 35.5 kg y exposure) no
$\beta\beta$-peak at $Q$=2039 keV was found.
The resulting half-live is
\begin{equation}\label{Hei-Mos}
T_{1/2}^{0\nu}(^{76}\mathrm{Ge})> 1.9 \times 10^{25} \, \mathrm{y}
\quad
(90 \% \mathrm{CL}),
\end{equation}
which implies that\footnote{Some participants of the
Heidelberg-Moscow experiment claimed\cite{Klapdor-Kleingrothaus:2006ff} the observation
of $0\nu\beta\beta$-decay of $^{76}\mathrm{Ge}$ with half-life
$T_{1/2}^{0\nu}(^{76}\mathrm{Ge}) = (2.23^{+0.44}_{-0.31}) \times 10^{25}$ y
(with 51.39 kg y exposure).
From this result the authors found
$|m_{\beta\beta}| = 0.32 \pm 0.03$ eV.
This claim will be checked by
the GERDA experiment\cite{Ur:2011zz}
using the same $0\nu\beta\beta$-decaying nucleus.}
\begin{equation}\label{Hei-Mos1}
|m_{\beta\beta}| \lesssim (0.22-0.64) \, \mathrm{eV}.
\end{equation}

In the cryogenic experiment Cuoricino\cite{1012.3266}
$\mathrm{TeO}_{2}$ bolometers were used,
with
a total mass of 11.34 kg of $^{130}\mathrm{Te}$.
The background was 0.17 counts/(kg~keV~y).
After a 19.75 kg y exposure the following lower bound was obtained:
\begin{equation}\label{Cuori}
T_{1/2}^{0\nu}(^{130}\mathrm{Te})> 2.8 \times 10^{24} \mathrm{y}\quad
(90 \% \mathrm{CL}),
\end{equation}
which corresponds to
\begin{equation}\label{Cuori1}
|m_{\beta\beta}| \lesssim (0.30-0.71) \, \mathrm{eV}.
\end{equation}

In the NEMO3 experiment\cite{hep-ex/0601021,Barabash:2010bd} the cylindrical source was
divided in sectors with enriched ${^{100}\mathrm{Mo}}$ (6914 g),
${^{82}\mathrm{Se}}$ (932 g) and other $\beta\beta$-decaying isotopes.
The two emitted electrons were detected in
drift cells and plastic scintillator.
No $0\nu\beta\beta$-decay was observed.
The half-life of $0\nu\beta\beta$-decay of $^{100}\mathrm{Mo}$ have been bounded by
\begin{equation}\label{Nemo}
T_{1/2}^{0\nu}(^{100}\mathrm{Mo})> 1.1 \times 10^{24} \, \mathrm{y}
\quad
(90 \% \mathrm{CL}).
\end{equation}
The corresponding limit for the effective Majorana mass is
\begin{equation}\label{Nemo1}
|m_{\beta\beta}| \lesssim (0.44-1.00) \, \mathrm{eV}.
\end{equation}

Several new experiments on the search for $0\nu\beta\beta$-decay
of different nuclei are currently running or in preparation.
In the following we discuss briefly some of them
(for more detailed presentations of future
experiments see Ref.\cite{GomezCadenas:2011it,Schwingenheuer:2012jt}).

In the GERDA experiment\cite{Ur:2011zz},
started in 2011, 18 kg of enriched germanium
crystals (with 86\% of the $\beta\beta$-decaying isotope $^{76} \mathrm{Ge}$) are
used. The expected background in the Phase-I of the experiment is $
10^{-2}$ counts/(kg~keV~y). After one year of running it is expected to reach
a sensitivity of
$T_{1/2}^{0\nu}(^{76}\mathrm{Ge})= 2.5 \times 10^{25}$ y,
which should allow to check the claim made in Ref.\cite{Klapdor-Kleingrothaus:2006ff}.

During the Phase-II of the GERDA experiment (expected to start in
2013), an array of enriched Germanium crystals (with 40 kg of $^{76}
\mathrm{Ge}$) will be cooled and shielded by liquid Argon of very
high radiopurity. A low background ($10^{-3}$ counts/(kg~keV~y)) is
expected. After 5 years of data taking, in the Phase-II of the
experiment a sensitivity of $T_{1/2}^{0\nu}(^{76}\mathrm{Ge})\simeq
1.9 \cdot 10^{26}$ y is expected. The corresponding
sensitivity to the effective Majorana mass is $
|m_{\beta\beta}|\simeq (7.3\cdot 10^{-2}-2.0\cdot 10^{-1})$ eV.

In the cryogenic CUORE experiment\cite{0912.0452}
$\mathrm{TeO_{2}}$ bolometers
are used both as source and as detector.
In the Phase-I of the experiment (started in the end of 2011)
the target mass is 10.8 kg of $^{130} \mathrm{Te}$. In the Phase-II (expected to start
in 2014) the target mass will be 206 kg of $^{130} \mathrm{Te}$. The expected background
in this phase will be $10^{-2}$ counts/(kg~keV~y). After 5 years of data taking
a sensitivity of $T_{1/2}^{0\nu}(^{130}\mathrm{Te})= 1.6 \cdot 10^{26}$ y
will be reached,
which corresponds to
$ |m_{\beta\beta}| \simeq (4.0-9.4) \cdot 10^{-2}$ eV.

In the KamLAND-Zen experiment\cite{1201.4664}, the $0\nu\beta\beta$-decay
of $^{136} \mathrm{Xe}$ will be studied.
In this experiment enriched $ \mathrm{Xe}$
(with 91\% of the $\beta\beta$-decaying isotope $^{136} \mathrm{Xe}$)
dissolved in liquid scintillator will be placed in a balloon
(3.4 m in diameter)
at the center of the KamLAND detector.
In the first phase of the experiment (started in 2011),
the source mass is 364 kg of $^{136} \mathrm{Xe}$.
In the second phase (scheduled for 2013)
910 kg of $^{136} \mathrm{Xe}$ will be utilized.
After 5 years of data taking it will be possible to reach a sensitivity
to $|m_{\beta\beta}|$
in the region of the inverted hierarchy
($|m_{\beta\beta}| \simeq 2.5 \cdot 10^{-2}$ eV).

In the running EXO experiment\cite{Gornea:2011zz} the decay
${}^{136}\mathrm{Xe}\to {}^{136}\mathrm{Ba}+e^{-}+e^{-}$
is searched for.
In
the first phase of the experiment (EXO-200) the mass of the fiducial volume is about 150 kg
of liquid Xenon enriched to 80.6\% in the $\beta\beta$-decaying isotope
$^{136} \mathrm{Xe}$.
After two years of data taking a sensitivity
$|m_{\beta\beta}| \simeq (8.7 \cdot 10^{-2} - 2.2 \cdot 10^{-1})$ eV is
planned to be achieved.
The full EXO experiment will consist of about 1 ton of enriched liquid Xenon.
With $ \mathrm{Ba}^{+}$ tagging,
a very low background of about $10^{-4}$ counts/(kg~keV~y) will be reached.
After 5 years of data taking,
it is expected to reach
a sensitivity of
$T_{1/2}^{0\nu}(^{136}\mathrm{Xe}) \simeq 10^{27}$ y,
which corresponds to
$|m_{\beta\beta}|\simeq (1.6-4.0)\cdot 10^{-2}$ eV.

\section{Conclusions}
\label{conclusions}

If massive neutrinos are Majorana particles,
neutrinoless double-beta decay of
$^{76}\mathrm{Ge}$,
$^{100}\mathrm{Mo}$,
$^{130}\mathrm{Te}$,
$^{136}\mathrm{Xe}$ and other even-even nuclei is allowed.
However, the
expected probability of $0\nu\beta\beta$-decay is extremely
small, because:

\begin{enumerate}

\item
It is a process of second order in the Fermi constant $G_{F}$.

\item
Since the Hamiltonian of weak interactions conserves helicity,
the amplitude of $0\nu\beta\beta$-decay is proportional to the
very small factor
\begin{equation}
\frac{m_{\beta\beta}}{\overline{q}^2},
\label{smallfactor}
\end{equation}
which comes from the neutrino propagator.
Here
$m_{\beta\beta}=\sum_{i}U^2_{ei}m_{i}$ is the effective Majorana
mass ($\lesssim 1$ eV) and $\overline{q}$ is the average neutrino momentum
($\sim 100$ MeV).

\end{enumerate}

The expected half-lives of $0\nu\beta\beta$-decays depend on the decaying nucleus
and are typically larger than $10^{24}-10^{25}$ years.
Therefore, the observation of this rare process is a real challenge.

The effective Majorana mass (and consequently the matrix element of the
process) depends on the character of the neutrino mass
spectrum.

In the case of a quasi-degenerate spectrum, the expected value of
$m_{\beta\beta}$ is relatively large. This case is partly excluded by
the data of the performed $0\nu\beta\beta$-decay experiments and
by cosmological
data
(see Fig.~\ref{bb0-plt}).
It will be further explored by GERDA, KamLAND-Zen, EXO, CUORE
and other experiments.

In order to reach the region of the inverted neutrino mass hierarchy,
with $10^{-2} \lesssim |m_{\beta\beta}| \lesssim 5 \cdot 10^{-2}$ eV,
the construction of large detectors ($\sim$ 1 ton) and about 5 years
of data taking will be required.

We considered here the $0\nu\beta\beta$-decay induced by the standard
mechanism of exchange of light Majorana neutrinos between $n$-$p$-$e^{-}$
vertices. From neutrino oscillation data it follows that if neutrino
with definite masses are Majorana particles this decay mechanism is realized
if there is no cancellation of the different mass contributions
(as shown in Fig.~\ref{bb0-plt},
cancellations can happen in the normal scheme).

As discussed in Section~\ref{seesaw},
the neutrino mass mechanism of $0\nu\beta\beta$-decay is predicted by the standard seesaw mechanism\cite{Minkowski:1977sc,GellMann-Ramond-Slansky-SeeSaw-1979,Yanagida-SeeSaw-1979,Mohapatra:1980ia}.
However, additional sources of violation of the
total lepton number $L$ are possible (see Ref.\cite{1103.6217} and references
therein). If $L$ is violated at the TeV scale these additional
mechanisms could give contributions to the matrix elements of the
$0\nu\beta\beta$-decay comparable with the contribution of the light
Majorana neutrino mass mechanism.

Let us consider as an example the violation of $L$ due to R-parity
violating interactions of SM and SUSY particles. In this case,
$0\nu\beta\beta$-decay is induced by the exchange of a heavy
Majorana SUSY neutralino. The product of $n$-$p$-$e^{-}$ vertices is
given by the factor
\begin{equation}\label{sUSY}
\left(\frac{G_{F}}{\sqrt{2}}\right)^2 \left(\frac{m^2_{W}}
{\Lambda^2}\right)^2\frac{1}{\Lambda},
\end{equation}
where $\Lambda$ characterizes the scale of the masses of SUSY particles and
$m_{W}$ is the mass of the $W$-boson. The factor (\ref{sUSY}) must be
compared with the corresponding factor
\begin{equation}\label{sUSY1}
\left(\frac{G_{F}}{\sqrt{2}}\right)^2\left(\frac{m_{\beta\beta}}{\overline{q}^2}\right),
\end{equation}
which appears in the case of the Majorana neutrino mass mechanism.
Taking into account that $\overline{q} \simeq 100$ MeV and assuming that
$|m_{\beta\beta}|\simeq 10^{-1}$ eV, we come to the conclusion that
Eqs.~(\ref{sUSY}) and (\ref{sUSY1}) are comparable if $\Lambda$ is of the order of a
few TeV.

If the $0\nu\beta\beta$-decay of different nuclei will be observed in
future experiments, it will be possible to probe the presence of
different mechanisms which can generate the process.

Finally, let us emphasize that
the search for $0\nu\beta\beta$-decay is a powerful practical
way to solve one of the most fundamental problem of modern neutrino
physics: {\em are neutrinos with definite masses $\nu_{i}$ truly
neutral Majorana particles or are they Dirac particles possessing a conserved
total lepton number?}

\section*{Acknowledgments}

We are grateful to F. \v{S}imkovic
who kindly provided us with Fig.~\ref{simkovic-nmes}.
We are also thankful to him for useful discussions.

\section*{Note Added}

After the completion of this review,
the EXO collaboration published in \texttt{arXiv:1205.5608}
the important first result of EXO-200.
With an exposure of 32.5 kg y, they obtained
$T_{1/2}^{0\nu}(^{136}\text{Xe}) > 1.6 \times 10^{25} \, \text{y}$ at 90\% CL, corresponding to
$|m_{\beta\beta}| \lesssim (0.14-0.38) \, \mathrm{eV}$.

\bibliography{bibtex/nu}

\begin{thebibliography}{10}

\bibitem{1002.3471}
Super-Kamiokande, R. Wendell et~al.,
Phys. Rev. D81 (2010) 092004, arXiv:1002.3471.

\bibitem{1109.0763}
SNO, B. Aharmim et~al.,
(2011), arXiv:1109.0763.

\bibitem{1009.4771}
KamLAND, A. Gando et~al.,
Phys. Rev. D83 (2011) 052002, arXiv:1009.4771.

\bibitem{hep-ex/0212007}
K2K, M.H. Ahn et~al.,
Phys. Rev. Lett. 90 (2003) 041801, hep-ex/0212007.

\bibitem{1103.0340}
MINOS, P. Adamson et~al.,
Phys. Rev. Lett. 106 (2011) 181801, arXiv:1103.0340.

\bibitem{Pontecorvo:1957cp}
B. Pontecorvo,
Sov. Phys. JETP 6 (1957) 429.

\bibitem{Pontecorvo:1958qd}
B. Pontecorvo,
Sov. Phys. JETP 7 (1958) 172.

\bibitem{Maki:1962mu}
Z. Maki, M. Nakagawa and S. Sakata,
Prog. Theor. Phys. 28 (1962) 870.

\bibitem{hep-ph/9812360}
S.M. Bilenky, C. Giunti and W. Grimus,
Prog. Part. Nucl. Phys. 43 (1999) 1, hep-ph/9812360.

\bibitem{Bilenky:1980cx}
S.M. Bilenky, J. Hosek and S.T. Petcov,
Phys. Lett. B94 (1980) 495.

\bibitem{Doi:1980yb}
M. Doi et~al.,
Phys. Lett. B102 (1981) 323.

\bibitem{Langacker:1986jv}
P. Langacker et~al.,
Nucl. Phys. B282 (1987) 589.

\bibitem{1001.0760}
C. Giunti,
Phys. Lett. B686 (2010) 41, arXiv:1001.0760.

\bibitem{1203.1669}
DAYA-BAY, F.P. An et~al.,
Phys. Rev. Lett. 108 (2012) 171803, arXiv:1203.1669.

\bibitem{1106.2822}
T2K, K. Abe et~al.,
Phys. Rev. Lett. 107 (2011) 041801, arXiv:1106.2822.

\bibitem{1108.0015}
MINOS, P. Adamson et~al.,
Phys. Rev. Lett. 107 (2011) 181802, arXiv:1108.0015.

\bibitem{1112.6353}
DOUBLE-CHOOZ, Y. Abe et~al.,
Phys. Rev. Lett. 108 (2012) 131801, arXiv:1112.6353.

\bibitem{0809.2936}
G.L. Fogli et~al.,
(2008), arXiv:0809.2936,
{NO-VE 2008, IV International Workshop on 'Neutrino Oscillations in
Venice' (Venice, Italy, April 15-18, 2008)}.

\bibitem{0804.3345}
A.B. Balantekin and D. Yilmaz,
J. Phys. G35 (2008) 075007, arXiv:0804.3345.

\bibitem{0806.2649}
G. Fogli et~al.,
Phys. Rev. Lett. 101 (2008) 141801, arXiv:0806.2649.

\bibitem{0810.5443}
H. Ge, C. Giunti and Q. Liu,
Phys. Rev. D80 (2009) 053009, arXiv:0810.5443.

\bibitem{1001.4524}
M. Gonzalez-Garcia, M. Maltoni and J. Salvado,
JHEP 04 (2010) 056, arXiv:1001.4524.

\bibitem{1203.1672}
Z. zhong Xing,
Chin. Phys. C36 (2012) 281, arXiv:1203.1672.

\bibitem{1003.5800}
M. Mezzetto and T. Schwetz,
J. Phys. G37 (2010) 103001, arXiv:1003.5800.

\bibitem{hep-ex/0104049}
LSND, A. Aguilar et~al.,
Phys. Rev. D64 (2001) 112007, hep-ex/0104049.

\bibitem{1007.1150}
MiniBooNE, A.A. Aguilar-Arevalo et~al.,
Phys. Rev. Lett. 105 (2010) 181801, arXiv:1007.1150.

\bibitem{1101.2755}
G. Mention et~al.,
Phys. Rev. D83 (2011) 073006, arXiv:1101.2755.

\bibitem{1101.2663}
T.A. Mueller et~al.,
Phys. Rev. C83 (2011) 054615, arXiv:1101.2663.

\bibitem{1106.0687}
P. Huber,
Phys. Rev. C84 (2011) 024617, arXiv:1106.0687.

\bibitem{1012.4356}
C. Giunti and M. Laveder,
Nucl. Phys. Proc. Suppl. 217 (2011) 193, arXiv:1012.4356,
{NOW 2010, 4-11 September 2010, Conca Specchiulla (Otranto, Lecce,
Italy)}.

\bibitem{hep-ex/0412056}
C. Kraus et~al.,
Eur. Phys. J. C40 (2005) 447, hep-ex/0412056.

\bibitem{1108.5034}
Troitsk, V. Aseev et~al.,
Phys. Rev. D84 (2011) 112003, arXiv:1108.5034.

\bibitem{Giunti-Kim-2007}
C. Giunti and C.W. Kim,
{Fundamentals of Neutrino Physics and Astrophysics} (Oxford
University Press, Oxford, UK, 2007),
{ISBN 978-0-19-850871-7}.

\bibitem{Minkowski:1977sc}
P. Minkowski,
Phys. Lett. B67 (1977) 421.

\bibitem{GellMann-Ramond-Slansky-SeeSaw-1979}
M. Gell-Mann, P. Ramond and R. Slansky,
(1979),
{In ``Supergravity'', p.~315, edited by F. van Nieuwenhuizen and D.
Freedman, North Holland, Amsterdam}.

\bibitem{Yanagida-SeeSaw-1979}
T. Yanagida,
(1979),
{Workshop on the Baryon Number of the Universe and Unified Theories,
Tsukuba, Japan, 13--14 Feb 1979}.

\bibitem{Mohapatra:1980ia}
R.N. Mohapatra and G. Senjanovic,
Phys. Rev. Lett. 44 (1980) 912.

\bibitem{Bilenky:2002aw}
S.M. Bilenky et~al.,
Phys. Rep. 379 (2003) 69, hep-ph/0211462.

\bibitem{Elliott:2002xe}
S.R. Elliott and P. Vogel,
Ann. Rev. Nucl. Part. Sci. 52 (2002) 115, hep-ph/0202264.

\bibitem{Elliott:2004hr}
S.R. Elliott and J. Engel,
J. Phys. G30 (2004) R183, hep-ph/0405078.

\bibitem{0708.1033}
I. Avignone, Frank~T., S.R. Elliott and J. Engel,
Rev. Mod. Phys. 80 (2008) 481, arXiv:0708.1033.

\bibitem{1001.1946}
S.M. Bilenky,
Lect. Notes Phys. 817 (2010) 139, arXiv:1001.1946.

\bibitem{1106.1334}
W. Rodejohann,
Int. J. Mod. Phys. E20 (2011) 1833, arXiv:1106.1334.

\bibitem{GomezCadenas:2011it}
J. Gomez-Cadenas et~al.,
Riv. Nuovo Cim. 35 (2012) 29, arXiv:1109.5515.

\bibitem{Schwingenheuer:2012jt}
B. Schwingenheuer,
(2012), arXiv:1201.4916,
{TAUP 2011}.

\bibitem{Weinberg:1979sa}
S. Weinberg,
Phys. Rev. Lett. 43 (1979) 1566.

\bibitem{hep-ph/9805219}
E. Ma,
Phys.Rev.Lett. 81 (1998) 1171, hep-ph/9805219.

\bibitem{0802.2962}
S. Davidson, E. Nardi and Y. Nir,
Phys. Rept. 466 (2008) 105, arXiv:0802.2962.

\bibitem{Doi:1985dx}
M. Doi, T. Kotani and E. Takasugi,
Prog. Theor. Phys. Suppl. 83 (1985) 1.

\bibitem{Bilenky:1987ty}
S.M. Bilenky and S.T. Petcov,
Rev. Mod. Phys. 59 (1987) 671.

\bibitem{Ryan-Okubo-NCS-2-234-1964}
C. Ryan and S. Okubo,
Nuovo Cimento Suppl. 2 (1964) 234.

\bibitem{Case:1957}
K.M. Case,
Phys. Rev. 107 (1957) 307.

\bibitem{hep-ph/9905509}
F. Simkovic et~al.,
Phys. Rev. C60 (1999) 055502, hep-ph/9905509.

\bibitem{0710.2055}
F. Simkovic et~al.,
Phys. Rev. C77 (2008) 045503, arXiv:0710.2055.

\bibitem{1007.0658}
S. Hannestad,
Prog. Part. Nucl. Phys. 65 2010 (2010) 185, arXiv:1007.0658.

\bibitem{hep-ph/9906525}
F. Vissani,
JHEP 06 (1999) 022, hep-ph/9906525.

\bibitem{hep-ph/0102265}
S.M. Bilenky, S. Pascoli and S.T. Petcov,
Phys. Rev. D64 (2001) 053010, hep-ph/0102265.

\bibitem{Schwetz:2011zk}
T. Schwetz, M. Tortola and J.W.F. Valle,
New J. Phys. 13 (2011) 109401, arXiv:1108.1376.

\bibitem{Bilenky:2010zza}
S. Bilenky,
{Introduction to the physics of massive and mixed neutrinos}
(Springer, 2010),
{Lecture Notes in Physics, Volume 817; ISBN 978-3-642-14042-6}.

\bibitem{hep-ex/0109033}
KATRIN, A. Osipowicz et~al.,
(2001), hep-ex/0109033.

\bibitem{Angrik:2005ep}
KATRIN, J. Angrik et~al.,
(2005).

\bibitem{1103.5083}
K.N. Abazajian et~al.,
Astropart. Phys. 35 (2011) 177, arXiv:1103.5083.

\bibitem{nucl-th/0305005}
V.A. Rodin et~al.,
Phys. Rev. C68 (2003) 044302, nucl-th/0305005.

\bibitem{0706.4304}
V.A. Rodin et~al.,
Nucl. Phys. A766 (2006) 107, arXiv:0706.4304.

\bibitem{nucl-th/0012010}
A. Bobyk, W.A. Kaminski and F. Simkovic,
Phys. Rev. C63 (2001) 051301, nucl-th/0012010.

\bibitem{Simkovic:2011zz}
F. Simkovic,
Phys.Part.Nucl. 42 (2011) 598.

\bibitem{1012.1783}
T.R. Rodriguez and G. Martinez-Pinedo,
Prog.Part.Nucl.Phys. 66 (2011) 436, arXiv:1012.1783.

\bibitem{1008.5260}
T.R. Rodriguez and G. Martinez-Pinedo,
Phys.Rev.Lett. 105 (2010) 252503, arXiv:1008.5260.

\bibitem{Rath:2010zz}
P.K. Rath et~al.,
Phys. Rev. C82 (2010) 064310.

\bibitem{Rath:2011zz}
P. Rath,
J.Phys.Conf.Ser. 322 (2011) 012019.

\bibitem{Barea:2009zz}
J. Barea and F. Iachello,
Phys. Rev. C79 (2009) 044301.

\bibitem{Iachello:2011zz}
F. Iachello and J. Barea,
Nucl.Phys.Proc.Suppl. 217 (2011) 5.

\bibitem{Iachello:2011zzb}
F. Iachello and J. Barea,
AIP Conf.Proc. 1355 (2011) 7.

\bibitem{Menendez:2011zza}
J. Menendez et~al.,
J.Phys.Conf.Ser. 312 (2011) 072005.

\bibitem{0801.3760}
J. Menendez et~al.,
Nucl.Phys. A818 (2009) 139, arXiv:0801.3760.

\bibitem{Simkovic-private-12}
F. \v{S}imkovic,
(2012),
{Private Communication}.

\bibitem{Klapdor-Kleingrothaus:2001yx}
H.V. Klapdor-Kleingrothaus et~al.,
Eur. Phys. J. A12 (2001) 147.

\bibitem{1012.3266}
CUORICINO, E. Andreotti et~al.,
Astropart. Phys. 34 (2011) 822, arXiv:1012.3266.

\bibitem{hep-ex/0601021}
NEMO, R. Arnold et~al.,
Nucl. Phys. A765 (2006) 483, hep-ex/0601021.

\bibitem{Barabash:2010bd}
NEMO, A. Barabash et~al.,
Phys.Atom.Nucl. 74 (2011) 312, arXiv:1002.2862,
{Fundamental Interactions Physics (ITEP, Moscow, November 23-27,
2009)}.

\bibitem{Klapdor-Kleingrothaus:2006ff}
H.V. Klapdor-Kleingrothaus and I.V. Krivosheina,
Mod. Phys. Lett. A21 (2006) 1547.

\bibitem{Ur:2011zz}
GERDA Collaboration, C. Ur,
Nucl.Phys.Proc.Suppl. 217 (2011) 38.

\bibitem{0912.0452}
F. Bellini et~al.,
Astropart. Phys. 33 (2010) 169, arXiv:0912.0452.

\bibitem{1201.4664}
KamLAND-Zen,
Phys. Rev. C85 (2012) 045504, arXiv:1201.4664.

\bibitem{Gornea:2011zz}
EXO Collaboration, R. Gornea,
J.Phys.Conf.Ser. 309 (2011) 012003.

\bibitem{1103.6217}
A. Ibarra, E. Molinaro and S.T. Petcov,
Phys. Rev. D84 (2011) 013005, arXiv:1103.6217.

\end{thebibliography}

\end{document}